\documentclass[twocolumn,epjc3]{svjour3} 
\usepackage{amssymb}
\usepackage{amsmath}
\usepackage{comment}
\usepackage{lineno}
\usepackage{comment}
\usepackage{overpic}

\RequirePackage{graphicx,subfig,multirow,ulem}

\DeclareGraphicsExtensions{.pdf}
\RequirePackage[numbers,sort&compress]{natbib}
\RequirePackage[colorlinks,citecolor=blue,urlcolor=blue,linkcolor=blue]{hyperref}

\newcommand{\DBD}{0$\nu\beta\beta$}

\newcommand{\TEO}{$\mathrm{TeO}_2$}
\newcommand{\TEHT}{$^{130}\mathrm{Te}$}

\newcommand{\CUORE}{CUORE}

\newcommand{\ckky}{\un{counts/(keV\,kg\,y)}}

\newcommand{\CERE}{Cherenkov}
\providecommand*{\un}[1]{\ensuremath{\mathrm{~#1}}}

\begin{document}       

\title{CALDER - Neutrinoless double-beta decay identification in TeO$_2$ bolometers with kinetic inductance detectors}

\author{
{E.S.~Battistelli}\thanksref{UNISAP}
\and
{F.~Bellini}\thanksref{UNISAP,INFN-RM1}
\and
{C.~Bucci}\thanksref{INFN-LNGS}
\and
{M.~Calvo}\thanksref{NEEL}
\and
{L.~Cardani}\thanksref{UNISAP,PRINCETON}
\and
{N.~Casali}\thanksref{UNISAP,INFN-RM1}
\and
{M.G.~Castellano}\thanksref{IFN}
\and
{I.~Colantoni}\thanksref{UNISAP}
\and
{A.~Coppolecchia}\thanksref{UNISAP}
\and
{C.~Cosmelli}\thanksref{UNISAP,INFN-RM1}
\and
{A.~Cruciani}\thanksref{UNISAP,INFN-RM1}
\and
{P.~de~Bernardis}\thanksref{UNISAP,INFN-RM1}
\and
{S.~Di~Domizio}\thanksref{UNIGE,INFN-GE}
\and
{A.~D'Addabbo}\thanksref{INFN-LNGS}
\and
{M.~Martinez}\thanksref{UNISAP,INFN-RM1}
\and
{S.~Masi}\thanksref{UNISAP,INFN-RM1}
\and
{L.~Pagnanini}\thanksref{INFN-LNGS,GSSI}
\and
{C.~Tomei}\thanksref{INFN-RM1}
\and
{M.~Vignati}\thanksref{INFN-RM1,UNISAP, e1}
}
\thankstext{e1}{e-mail: marco.vignati@roma1.infn.it}

\institute{{Dipartimento di Fisica, Sapienza Universit\`{a} di Roma, Roma - Italy}\label{UNISAP}
\and
{INFN Sezione di Roma, Roma - Italy}\label{INFN-RM1}
\and
{INFN  Laboratori Nazionali del Gran Sasso, Assergi (AQ) - Italy}\label{INFN-LNGS}
\and
{Institut N\'eel - CNRS, Saint-Martin-d'H\'eres - France}\label{NEEL}
\and
{Physics Department, Princeton University, Princeton, NJ - USA}\label{PRINCETON}
\and
{Istituto di Fotonica e Nanotecnologie - CNR, Roma - Italy}\label{IFN}
\and
{Dipartimento di Fisica, Universit\`{a} di Genova, Genova - Italy}\label{UNIGE}
\and
{INFN  Sezione di Genova, Genova - Italy}\label{INFN-GE}
\and
{INFN Gran Sasso Science Institute, L'Aquila - Italy}\label{GSSI} 
}

\maketitle

\begin{abstract}
Next-generation experiments searching for neutrinoless double-beta decay must be sensitive
to a  Majorana neutrino mass as low as 10\un{meV}. \CUORE, an array of 988 TeO$_2$ bolometers being commissioned at Laboratori Nazionali del Gran Sasso in Italy, features an expected sensitivity of 50-130\un{meV} at 90\%~C.L , that can be improved by removing the background from $\alpha$ radioactivity. This is possible if, in coincidence with the heat release in a bolometer, the \CERE\ light emitted by the $\beta$ signal is detected. The  amount of light detected is so far limited to only 100\un{eV},  requiring low-noise cryogenic light detectors. The CALDER  project (Cryogenic wide-Area Light Detectors with Excellent Resolution) aims at developing a small prototype experiment consisting of TeO$_2$ bolometers coupled to new light detectors based on kinetic inductance detectors. The R\&D is focused on the light detectors that could be implemented in a next-generation neutrinoless double-beta decay experiment.
\end{abstract}

\PACS{23.40.-s, 07.57.Kp, 29.40.Ka, 84.40.Dc}
\keywords{Neutrinoless double beta decay, Bolometer, \CERE\ detector, Kinetic Inductance Detector}


\section{Introduction}

Bolometers proved to be good detectors to search for neutrinoless double-beta decay (\DBD), thanks to the possibility of studying different isotopes,
the excellent energy resolution, and the low background they can achieve~\cite{Artusa:2014wnl}.
The \CUORE\ experiment~\cite{Artusa:2014lgv} will search for the \DBD\ of \TEHT\ using an array of 988~\TEO\ bolometers
operated at a temperature around 10\un{mK}. 
Each bolometer weighs 750\un{g}, for a total active mass of 741\un{kg},  206\un{kg} of which are \TEHT\ (34.2\% natural abundance in tellurium~\cite{Fehr200483}). The energy resolution and the background at the $Q$-value of the decay ($Q_{\beta\beta}$=2528\un{keV}~\cite{Redshaw:2009zz}), are expected to be 5\un{keV~FWHM} and $10^{-2} \ckky$, respectively~\cite{Aguirre:2014lua}. 
\CUORE\ is in construction at Laboratori Nazionali del Gran Sasso (LNGS) in Italy, and is expected to start operations within one year.
The 90\% C.L. sensitivity of \CUORE\ to the \DBD\ half-life is predicted to be  $10^{26}$ years in 5 years of data taking~\cite{Aguirre:2014lua}.
 This corresponds to an effective neutrino Majorana mass that ranges, depending on the choice of the nuclear matrix elements, from $50$ to $130\un{meV}$,  values that are quite far from covering the entire interval of masses corresponding to the inverted hierarchy scenario, that ranges from $10$ to $50\un{meV}$~\cite{Bilenky:2012qi}. 
 
 The sensitivity of CUORE is limited by two factors, the amount of isotope and the background level, which is expected to be dominated by $\alpha$ radioactivity. To overcome these limits, a new experiment to be run after CUORE is being designed, CUPID~\cite{CUPIDsci,CUPIDRD}. 
  The technology of CUPID is not yet defined and is open to many alternatives. One of the main options is to use scintillating crystals enriched in high-$Q_{\beta\beta}$ isotopes, such as Zn$^{82}$Se~\cite{Beeman:2013sba}, Zn$^{100}$MoO$_4$~\cite{Beeman:2012ci} or $^{116}$CdWO$_4$~\cite{Arnaboldi:2010tt}. The scintillation light is detected in coincidence with the heat release in the bolometer and is used to discriminate the $\beta$ signal from the $\alpha$s, exploiting the dependence of the light yield on the ionization power of  incident particles.  The other option  is to use again  \TEO\ bolometers, enriched in \TEHT, profiting from the experience acquired within \CUORE\ and from the much cheaper enrichment. Unfortunately \TEO\ is not a scintillator, and other means to remove the $\alpha$ background must be found.
 
A promising alternative consists in detecting the small amount of  \CERE\ light that is emitted by  particles absorbed in the \TEO\ crystal. At the energy scale of interest for \DBD, $\beta$ particles  are above threshold for \CERE\ emission, while $\alpha$s are not~\cite{TabarellideFatis:2009zz}. In a recent work~\cite{Casali:2014vvt}, the \CERE\ light emitted by $\beta$ particles in a \CUORE\ crystal was measured and found  to be
100~eV at $Q_{\beta\beta}$. The light detector consisted in a germanium disk read by a Neutron Transmutation Doped (NTD) germanium thermistor~\cite{Beeman:2013zva}, which was originally developed to detect the much larger amount of light emitted by scintillating crystals (several keV). The light detector noise amounted to 70\un{eV~RMS}, a level too high to allow an event by event discrimination.
It was computed that, to reject the $\alpha$ background, one needs light detectors featuring a  noise smaller than 20~eV~RMS.

Many light detector technologies are being proposed for CUPID (see~\cite{CUPIDRD} and references therein): 
1) CRESST-like detectors based on superconducting Transition Edge Sensors (TES), 
2) NTD-based detectors exploiting the Neganov-Luke effect, 
3) Metallic Magnetic Calorimeters (MMC) and 4) Kinetic Inductance Detectors (KIDs), the detectors proposed in this paper.
It has to be stressed that high-sensitivity light detectors, though not strictly required, could also be applied to scintillating crystals. In the case of ZnSe they would allow to discriminate
nuclear recoils from electron recoils at low energies and thus enable the search for direct Dark Matter interactions~\cite{Beeman:2013vda}.
The technology chosen for CUPID, aside the achievement of the resolution goal, must prove to be reproducible and scalable to a thousand light detectors, and easily implementable in the CUORE infrastructure~\cite{CUPIDsci}. 

Among the different technologies, TESs already proved to feature the required energy resolution~\cite{Schaffner:2014caa,Willers:2014eoa}, however
they need an extra R\&D for CUPID. Citing~\cite{CUPIDRD}: ``The most important issue is the reproducibility of the technology (e.g. uniformity of transition temperature across many channels) at temperatures of order 10 mK and the cost and effort required for a construction of a large quantity of high-quality detectors. Additional aspects, such as multiplexing of the detector signals to reduce the wiring complexity and the heat load,
 would be useful: solutions already exist in the astrophysics community.'' For this reason, we decided to use Kinetic Inductance Detectors~\cite{Day:2003fk} (KIDs), that offer two advantages with respect to TES-based devices. 
 
First, KID performances do not depend critically on the working temperature, since it is sufficient that the temperature is well below the critical temperature of the superconductor. Second, the readout electronics is quite simple: KIDs are naturally multiplexable and their electronics is operated at room temperature, exception made for a low noise cryogenic amplifier.  KIDs recently demonstrated to be a valid alternative to TESs in astrophysical applications~\cite{Monfardini2011,Mazin:2013wvi}: they feature similar sensitivity, but they are  easier to operate and can be scaled to a large number of detectors.

Light detectors  with active areas as large as the face of CUORE crystals (5x5\un{cm^2}) based on KIDs do not exist yet. The goal of the CALDER
project is to realize such detectors and test their performances on a  prototype experiment made of a few \TEO\ bolometers.

\section{Kinetic Inductance Detectors}\label{sec:kids}
KIDs base their working principle on  the kinetic inductance.
In superconducting materials the Cooper pairs, characterized by a binding energy smaller than 1$\,$meV, move through the lattice without scattering. Nevertheless the complex impedance is non-zero. If an RF e-m field is applied, the pairs change continuously their velocity, and their inertia, due to the stored kinetic and magnetic energy, generates an impedance. The kinetic component of the impedance depends on the density of Cooper pairs, which can be modified by an energy release able to break them into quasiparticles (i.e. particles that, for the sake of simplicity, can be considered as free electrons). If the superconductor is inserted in a resonant RLC circuit with high quality factor ($Q>10^3$), the density variation of quasiparticles produces changes in the transfer function, both in phase and amplitude. 

The signal is obtained by exciting the circuit at the resonant frequency, and by measuring the phase (inductance) and amplitude (resistance) variations induced by energy releases. Many KIDs  can be coupled to the same feedline, and can be multiplexed by making them resonate at slightly different frequencies. The resonant frequency of each resonator ($f_0=1/2\pi\sqrt{LC}$) can be easily varied by slightly changing the layout of the capacitor and/or inductor of the circuit. 

The voltage transmitted by a KID with resonant frequency $f_0$ can be written as a function of frequency $f$ as~\cite{zmu_annrev2012}:
\begin{equation}
S_{21} = 1 - \frac{Q/Q_c}{1+2\jmath Q \frac{f-f_0}{f_0}}
\end{equation}
where the quality factor $Q$  is a function of the internal and coupling quality factors:
${1 \over Q} = {1\over Q_i} + {1\over Q_c }$.
The coupling quality factor $Q_c$ depends on the device geometry and line impedance.
The internal quality factor  $Q_i$ is a combination of the  quality factor due to the thermally generated quasiparticles and the quality factor due to other loss mechanisms 
such as film and substrate impurities and irradiation.

 The signal induced by an excess number of quasiparticles $N_{qp}(t)$ is~\cite{gaoPhD}:
 \begin{equation}\label{eq:delta_s21}
 \Delta S_{21}(t) = 
 \frac{\alpha Q^2}{Q_c} 
 \frac{N_{qp}(t)}{2 N_0 V\Delta}
 \left[S_1(f_0,T) -\jmath S_2(f_0,T)\right]
 \end{equation}
where $T$ is the working temperature, $V$ the resonator volume, $2\Delta$ is the binding energy of Cooper pairs, $\alpha$ is the
fraction of kinetic inductance over the total inductance, $N_0$ is  the single-spin density of states at the Fermi energy, $S_1(f_0,T)$ a dimensionless factor around 1 describing the amplitude response  
and $S_2(f_0,T)$ a factor greater than 1 describing phase response.
Since the phase response is larger, it is often the only one considered. 

The number of created quasiparticles is proportional to the particle energy and decays with
the recombination time  $\tau_{qp}$: 
\begin{equation}
N_{qp}(t) = \eta \frac{E}{\Delta} e^{-t/\tau_{qp}}
\end{equation}
where $\eta$ is the energy conversion efficiency to quasiparticles. The rise time of the signal is limited by the resonator ring time ($\tau_r = Q/\pi f_0$), which in present detectors is much smaller than $\tau_{qp}$.

The ultimate source of noise in KIDs is due to the generation-recombination effect originating from the dynamic equilibrium between Cooper pairs and thermally created quasiparticles. Since the number of quasiparticles decreases exponentially with  temperature, KIDs are operated well below the critical temperature ($T<T_c/6$). The noise, however, is in practice limited by amplifier noise or by noise induced by two level systems (TLS) in the substrate (see for example~\cite{gaoPhD}). TLS noise is however expected to dominate at frequencies lower than the signal bandwidth we are interested in~\cite{moore2}, therefore it will not be considered here. 

Following Refs.~\cite{zmu_annrev2012,moseley:1257}, we can compute the energy resolution for the phase readout in the amplifier-limited noise case:
\begin{equation}\label{eq:resosingle}
\sigma_E^K= \frac{\Delta}{\eta}  \frac{ 2 N_0 V \Delta}{\alpha Q S_2(f_0,T)} \sqrt{\frac{ k_B T_N}{ P_f \tau_{qp}}}  
\end{equation}
where $T_N$ is the amplifier noise temperature, $P_f$ is the KID readout power and, in a conservative scenario, we assume to work with overcoupled resonators ($Q_c\ll Q_i$). It is therefore clear that, to boost the energy resolution, one has to design KIDs with high $\eta$, $\alpha$, $Q$ and $\tau_{qp}$. Rising $P_f$ is also needed to overcome the amplifier noise but, since a fraction of the power is absorbed by the resonator, it also induces an unwanted lowering of $Q_i$ (and  therefore $Q$) and $\tau_{qp}$.

By using typical values for aluminum resonators ($\alpha=10\%,Q=10^{4}, \tau_{qp}=100\un{\mu s},S_2=3, \Delta = 230\un{\mu eV}, N_0=1.72\times 10^{10}\un{\mu m^{-3} eV^{-1}}$), $P_f = -70\un{dBm}$, $T_N=5\un{K}$ (see more in Sec.~\ref{sec:readout}) and a resonator volume 
$V=1\un{mm^2}\times40\un{nm}$, the expected energy resolution is $\sigma_E^K \sim 2\un{eV}/\eta$, where $\eta=0.57$ for  absorbed optical photons~\cite{Day:2003fk}.

\section{Phonon-mediated light detectors}

KIDs, in the so-called ``lumped element'' configuration we chose to implement~\cite{Doyle2008}, are composed of two different parts, a meander acting as inductor and a set of interdigitated lines acting as capacitor.
Dimensions are chosen smaller than the wavelength of the excitation signal, so that the current in the inductor is uniform and the signal does not depend on the position of the energy release.

KIDs can reach a maximum size of a few \un{mm^2}, which is very small considering that we need an active area of 5x5\un{cm^2}. Increasing the size of the individual resonator would lower the resonant frequency below the optimal range (1-4\un{GHz}), thus loosing the possibility to use already available electronics and reducing the number of resonators that can be coupled to the same line. 
On the other hand, covering 
the entire area with KIDs is not possible. It would require thousands of pixels per light detector, a number that would be impossible to scale to thousands light detectors. Another problem is the optical coupling between the pixel and the photons. Obtaining at the same time large and fully active areas is difficult, even if the KIDs themselves act as absorbers or some kind of collector is applied.

The solution we opted for consists in 
using the  silicon substrate on which the KIDs are deposited  to perform an indirect observation, as in CRESST light detectors~\cite{Angloher:2011uu}. 
Photons impinging on the back side of the chip can produce athermal or ballistic phonons that  scatter through
the substrate and  reach the KID on the opposite surface, where they break Cooper pairs and generate a signal.
The advantage of using phonons as mediators is that they can propagate along distances longer than a centimeter allowing to sample an area larger than the resonator itself.  
 The disadvantage is that the efficiency is lower than for direct absorption, since phonons in the substrate
can be absorbed by the surfaces of the substrate itself or by the detector supports or they can decay to sub-gap phonons before reaching the superconductor. 
 
To compensate the efficiency loss with respect to direct absorption, a few KIDs per light detector will be needed. 
The expected energy resolution combining $N_K$ KIDs deposited on the same substrate is simply:
\begin{equation}
\sigma_E = \sigma_E^K\sqrt{N_K} 
\end{equation}
where the efficiency $\eta$ in $\sigma_E^K$ (Eq.~\ref{eq:resosingle}) is  now the overall efficiency in converting energy deposited in the substrate to quasiparticles in the sensors.

Phonon-mediated KIDs were already proposed by Swenson et al.~\cite{swenson} and  by Moore et al.~\cite{moore2}  to measure  simoultaneously energy and  position of  impinging X-rays. Moore et al. realized a detector consisting of a silicon absorber of $2\times2\un{cm^2}\times1\un{mm}$  sampled  by $20$ aluminum resonators  coupled to the same feedline with an active volume of $1\un{mm^2}\times25\un{nm}$ each. The quality factor of the resonators was limited to $Q=10^4$ by design, so as to have a response sufficiently fast ($\tau_r \sim 1\un{\mu s}$) to discriminate the arrival time of phonons and infer the position of the interaction. The measured efficiency was $\eta=0.07$, the fraction of kinetic inductance was $\alpha=0.075$, while the recombination time was limited to $\tau_{qp} = 13\un{\mu s}$. 
The energy resolution amounted to 550\un{eV} on 30\un{keV} X-rays and to 380\un{eV} on the detector baseline. 

Since we are not interested in the position reconstruction, we can design detectors with a smaller number of pixels ($<10$), compensating the possible efficiency loss with a higher $Q$ (see Eq.~\ref{eq:resosingle}). 
The energy resolution can be further improved by using superconductors with higher kinetic inductance than aluminum. TiN and multilayer Ti/TiN films, for example, proved to feature a  kinetic inductance which is a factor 10-30 higher than aluminum films~\cite{leduc:102509,vissers:232603}.

\section{Light detector design and implementation}
Detectors are designed and simulated using SONNET, a 3D planar software
for radiofrequency analysis~\cite{sonnet}. This software enables the study of
the device features ($f_{0}, \alpha,Q_{c}$)
by simulating  the circuit properties starting from the chip geometry. 
Since the readout performances are optimal between 1-4\un{GHz},
the geometric inductance and
capacitance are designed to obtain a resonant frequency $f_0$ around 2.5\un{GHz}. 

\begin{figure}
\begin{center}
\includegraphics[width=0.48\textwidth]{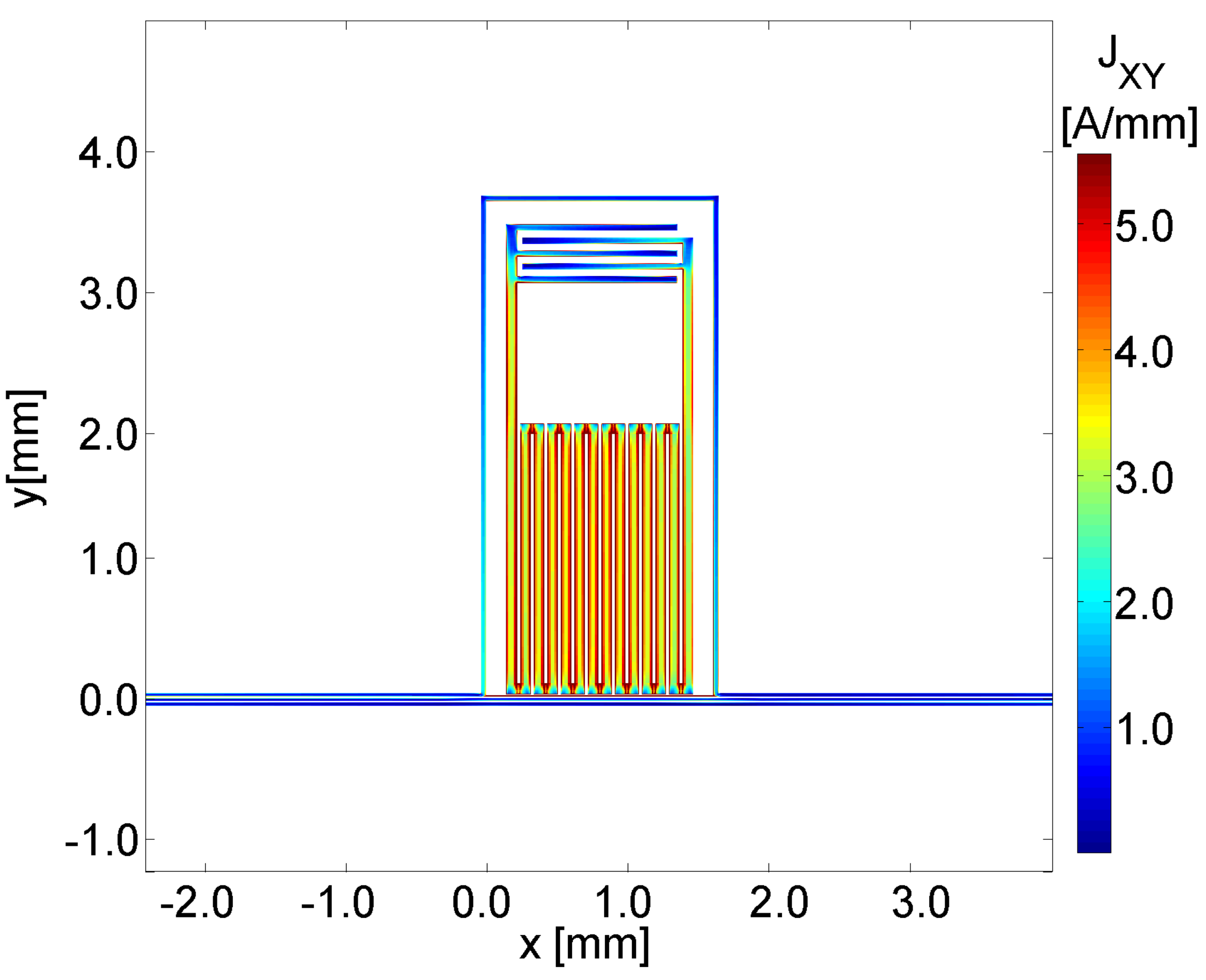}
\caption{SONNET simulation of the on-resonance current flowing in a single pixel. The pixel is composed by an inductive meander, which is the active area,  and by an interdigited capacitor to enlarge the resonant wavelength and make uniform the current across the inductor (see description in the text). }
\label{fig:pixel}
\end{center}
\end{figure}

The  design aims at obtaining at the same time a
large active area, a high resonant frequency  and a high fraction of
kinetic inductance $\alpha$.  The prototype single pixel we present
is $1.4\times3.5\un{mm^2}$ wide and, as shown in Fig.~\ref{fig:pixel},
consists of an inductive meander (14 connected strips of $80\un{\mu
m}\times2\un{mm}$), representing the active area, and a capacitor
(5 interdigitated fingers of $1.2\un{mm}\times 50\un{\mu m}$). To reduce the geometric
inductance in favor of the kinetic inductance the spacing between
meanders is minimized ($20\un{\mu m}$) resulting in an estimate of $\alpha=5\%$ for a 40\un{nm}  Al film. 
The capacitance is chosen to keep 
the current flowing through the meander uniform within a factor 20\%. The width of the fingers and their
distance ($50\un{\mu m}$) are designed to decrease the local electric
field and, consequently, TLS noise~\cite{noroozian2009}.

\begin{figure}
\begin{center}
\includegraphics[width=0.5\textwidth]{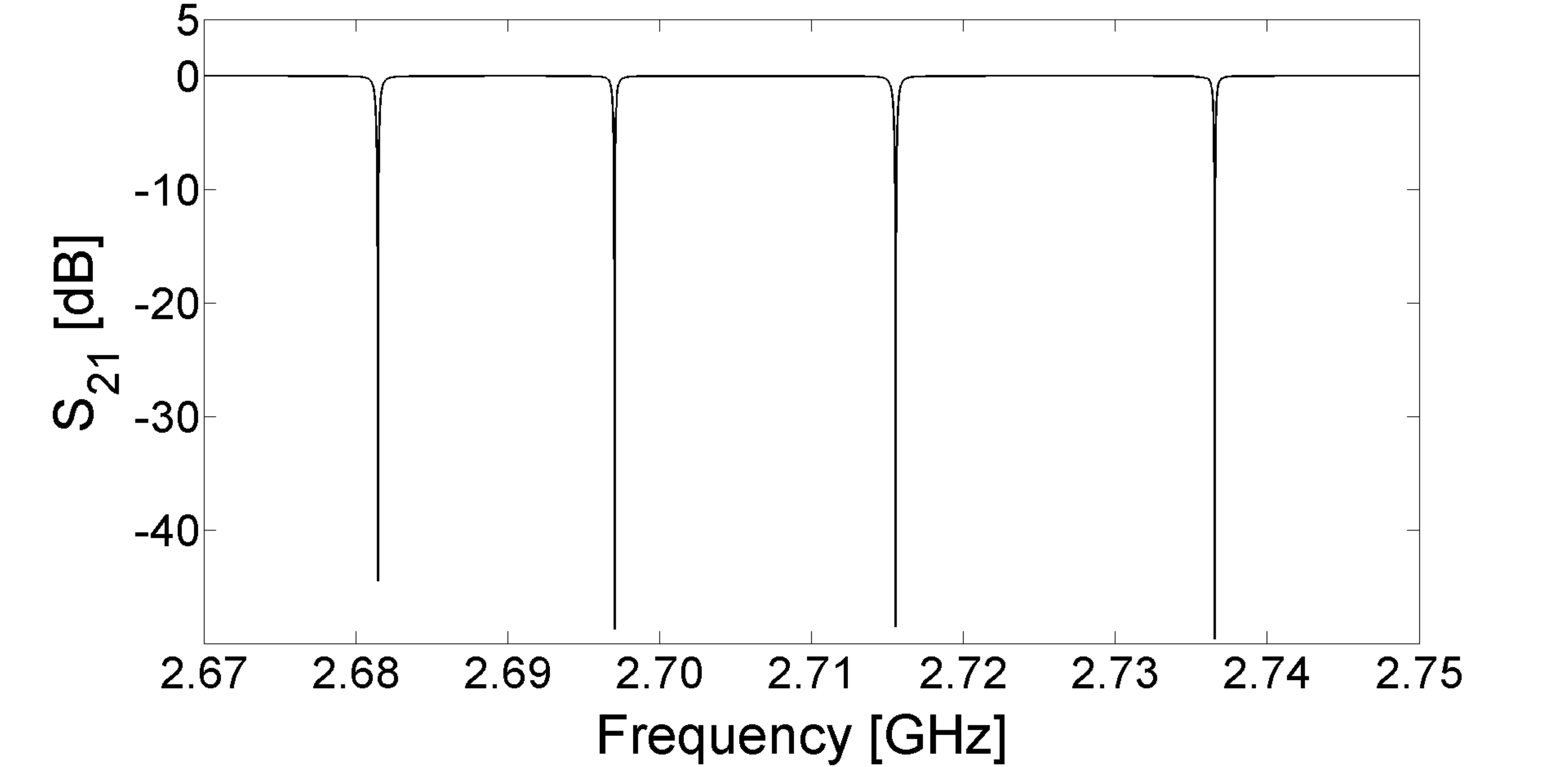}
\caption{Sonnet simulation of the frequency response of a 4-pixel array. The 4 pixels are coupled to the same feedline and spaced in frequency by changing the length of the last capacitor finger.}
\label{fig:4pix}
\end{center}
\end{figure}

The pixel is inductively coupled to a Coplanar Waveguide (CPW) and surrounded by a ring ground to reduce the cross-talk among adiacent pixels. By
changing the distance between the pixel and the CPW we determine
$Q_c$, which in the first prototypes is in the range
$10^{4}-10^{5}$.  The ground plane width of the CPW is minimised in
order to reduce the probability of absorbing phonons, since this region is inactive for detection.
For a small number of pixels (up to $\sim 10$), SONNET allows to simulate
the whole array, so that cross-talk effects can be studied and minimized.
In Fig.~\ref{fig:4pix} we show the simulated frequency response of a prototype array of 4 pixels.
The frequency spacing of the 4 resonators is simply realized by reducing the length of the last finger of the capacitance.

The detectors are produced in a ISO5/ISO6 clean room at IFN-CNR, using a lift-off technique. Devices are fabricated on high-quality (FZ method) $3"\times300\un{\mu m}$ intrinsic Si(100) wafers, with high resistivity ($\rho>10\un{k\Omega\cdot cm}$) and double side polished.
Wafers are coated with 300\un{nm}  positive electronic resist (PMMA 6\%), and exposed to a 100\un{keV}  electron beam  for patterning. This  technology allows to easily change the KIDs geometry run by run, just at cost of exposition time. After exposure and development, a gentle cleaning in Oxygen plasma is performed to produce an undercut in the PMMA profile and native oxide is removed from the silicon areas by dipping the wafer in dilute HF at 2\% for 10\un{sec}.
The patterned substrate is then covered with a 40\un{nm} aluminum layer deposited by electron gun evaporation, in a HV chamber, at a rate of 0.8\un{nm/s}.  After the Al deposition, the wafer is immersed in hot acetone to achieve lift-off: the e-resist under the Al film is dissolved taking the excess film with it. 

The wafer is finally cut into single 20x20$\,$mm$^2$ chips and the Al thickness is measured by profilometer. The chip is mounted in a copper holder, fixed by PTFE supports with a contact area of few $\rm{mm^2}$ to prevent trapping of the phonon signal, and connected to SMA read-out by ultrasonic Al wire bonding. 

In the near future, TiN and multilayer Ti/TiN films will be also deposited, by means of reactive magnetron sputtering and the pattern will be transferred with subtractive method (dry or wet etch).

\section{Readout}\label{sec:readout}
The simplicity of the signal readout is among the most striking advantages of kinetic inductance detectors.
All the complexity of the frequency-multiplexed readout is handled by  room temperature digital electronics, and only a limited number of components are located at low temperature inside the cryostat.
The only limiting factors for the number of resonators that can be coupled to a single feedline come from the bandwidth and the dynamics of the DAC and ADC that are used for the signal excitation and readout, and by the minimum spacing (either in frequency and in physical distance) among adjacent resonators that results in the maximum tolerable cross talk level.

\begin{figure}
\begin{center}
\resizebox{0.35\textwidth}{!} {
  \includegraphics{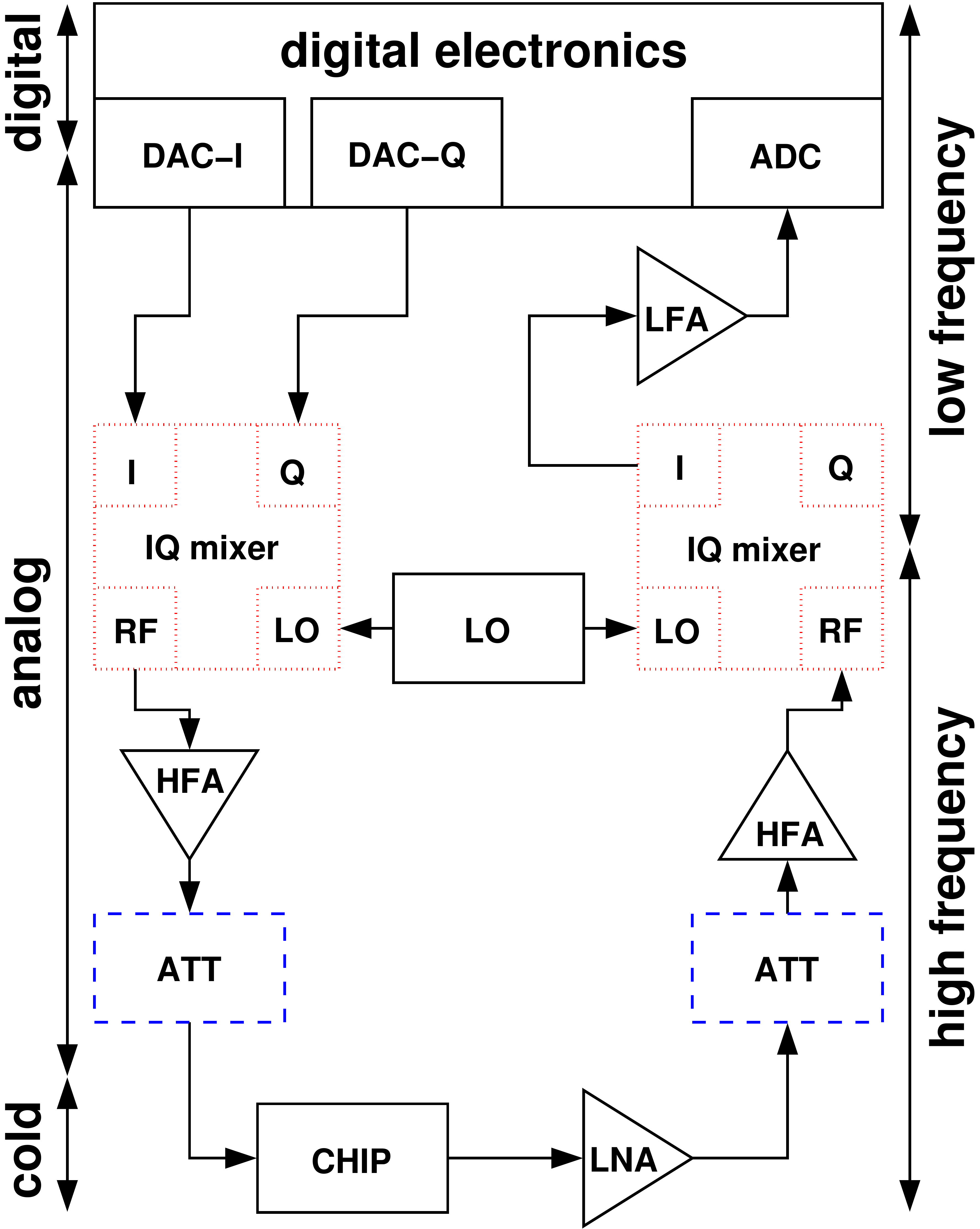}
}
\end{center}
\caption{CALDER readout scheme. The I and Q excitation signals are generated by the DAC with a bandwidth of about 100$\,$MHz and are sent to an I-Q mixer, that also takes as input a $\sim$2.5$\,$GHz sinusoidal signal from a local oscillator (LO).
  The up-converted output (RF) of the I-Q mixer is sent to a high frequency room temperature amplifier (HFA) and then to a remotely controllable variable attenuator (ATT).
  Then the signal enters the cryostat where it goes through the detector (CHIP) and a low noise cryogenic amplifier (LNA).
  Outside the cryostat, the signal passes through another variable attenuator and a high frequency amplifier, and is then down-converted by another I-Q mixer.
  Finally, the signal passes through a low frequency amplifier (LFA), used to match the signal level to the ADC input.}
\label{fig:readout}
\end{figure}

The readout system follows the scheme described e.g. in Refs.\cite{McHugh:2012dy, Bourrion:2011gi}.
It can be divided conceptually in three blocks (see Fig.~\ref{fig:readout}): the cold electronics, the room temperature analog electronics and the room temperature digital electronics.

The cold electronics includes a limited number of components.
These are the feedline, a low noise, wide band cryogenic amplifier and some passive components such as DC-blocks and attenuators.
The cryogenic amplifier is based on a SiGe high electron mobility transistor (HEMT).
It is located in the dilution refrigerator past the detectors, and is thermally connected to the 4\un{K} plate.
The typical values for the gain and the noise temperature are of 40$\,$dB and about 5$\,$K, respectively, and they are almost constant in a wide frequency band extending from 1 to 4$\,$GHz.
With a careful design, the noise coming  from components of the readout system can be made negligible with respect to that of the cryogenic amplifier.

The room temperature electronics has the role of matching the signal levels and the frequency bandwidth of the DAC and the ADC to those of the cold part of the readout chain, while keeping the noise level sufficiently low.
In fact, the resonant frequency of  KIDs is typically in the few GHz range, and there are no DAC and ADC available at present that feature at the same time such a large bandwidth and a proper resolution.
The ADC and the DAC work best with typical powers of about 10$\,$dBm and a bandwidth of about 100$\,$MHz, while the CALDER sensors must be excited with powers in the range from -100 to -50\un{dBm} at frequencies around 2.5$\,$GHz.
The excitation signal at the output of the DAC is therefore fed to a mixer that also takes as input a $\sim$2.5$\,$GHz sinusoidal signal and up-converts the excitation signal to the frequency range corresponding to the sensor resonances.
At the cryostat output, the signal is down-converted to low frequency by means of another mixer and is then fed to the ADC.
The $\sim$2.5$\,$GHz sinusoidal signal used as input of the two mixers is generated by a commercial local oscillator (LO) with an extremely low phase noise ($\sim -140\un{dBc/Hz}$ at 10\un{kHz} from the carrier).
Besides performing the frequency up- and down-conversion of the signal, the room temperature analog electronics also includes  attenuators, DC-blocks and amplifiers, both in the low and high frequency parts.
These are used to control the excitation power that is sent to the sensors, and to match the signal levels to the specifications of the various components of the readout system, so that they work at their optimal conditions and the noise is minimized.

The room temperature digital electronics can be conceptually sketched as an FPGA interfaced on one side to the ADC and the DAC, and to a computer on the other side.
In our setup we are currently using the NIXA hardware implementation~\cite{Bourrion:2011gi} with the firmware described in~\cite{Bourrion:2013ifa}, but we are also considering the ROACH-2 system as an alternative~\cite{McHugh:2012dy}.
In the following we give a general description of the digital electronics system that applies to both the implementations.
The FPGA firmware is composed by a core that is responsible for generating the excitation signal for the DAC and then demodulating it after it is read back by the ADC, and by a digital processing unit that essentially has the purpose of running a low threshold trigger algorithm.
The generation of the excitation signal is one of the most resource consuming parts of the firmware.
For a KID resonating at frequency $f_0$, the excitation is composed of an {\it in phase} (I) sine wave at frequency $f_{base}=f_0-f_{LO}$, where $f_{LO}$ is the frequency of the LO, and by a {\it quadrature phase} (Q) sine wave with the same frequency and a $\pi$/2 phase shift with respect to the first one.
The overall I(Q) comb excitation signal is then built as the sum of all the I(Q) signals of the resonators on the feedline, and is then sent to the DAC.
When the comb signal is read back by the ADC, the component corresponding to each resonator is demodulated with a copy of its own excitation signal, thus resulting in a pair of low frequency (I, Q) signals for each resonator.
The signal bandwidth and the sampling frequency of the demodulated signals depend on the resonators, but they are  typically of order 100\un{kHz} and 1\un{MHz}, respectively.
The demodulated I and Q signals are further manipulated with digital filters, and then the trigger algorithms are run.
As discussed in Sec.~\ref{sec:kids}, the detector response is more sensitive in the phase direction, therefore it would be desirable to trigger on this parameter.
However, because the calculation of the phase as $\arctan$(Q/I) requires too intensive FPGA resources, an artifact is used that consists in mapping the original (I, Q) signals into rotated (I$_{ROT}$, Q$_{ROT}$) values, and the trigger is run on them.
The rotation angles must be pre-calculated for each resonator during the detector characterization.
A proper choice of their values can make one of I$_{ROT}$ or Q$_{ROT}$ proportional to the signal phase, while the other almost insensitive to changes in the kinetic inductance.
A window of configurable length of the sensor waveform around the position of each trigger is streamed via network to a data acquisition computer, where a prompt analysis is performed, and then the data are saved to disk for offline analysis.

To operate the CALDER sensors as auxiliary light detectors for a large array of macrobolometers, a few more steps will be needed.
First, it is unlikely that the readout of all the CALDER light detectors will be possible with a single feedline and a single digital electronics board.
Therefore, a method for the data synchronization, merging and streaming from multiple KID readout systems must be foreseen.
Second, an integration of the data acquisition systems of the light detectors and of bolometers will be necessary.
The particle discrimination capability provided by the CALDER detectors would require in fact to perform a synchronization of the KIDs with the bolometers.
Moreover, a bolometer trigger should also cause a KID event to be saved, and vice-versa.
In this sense some care will be needed to deal with the very different evolution time scales of the signals from these two detectors.
Signals from large mass bolometers have typical rise times of tens of milliseconds, while the KID signals have typical rise times of some microseconds.

\section{Cryogenic tests}

Kinetic inductance detectors reach their best performances at
temperatures lower than about 1/6 of the critical temperature of the
superconductor, demanding for operation temperatures of 50-100\un{mK}. 

In this first phase of the project the
light detectors are being tested in our laboratory in Rome, which is equipped
with a wet $^3$He/$^4$He dilution refrigerator with base temperature of 10\un{mK} and
radiative shields at 600\un{mK} and 4.2\un{K}. There are 4 RF coaxial cables
installed in the cryostat, allowing the independent test of 2 devices
simultaneously.

The detectors can be  illuminated 
by a multimode optical fiber, coupled to a warm LED ($\lambda=
400\un{nm}$) able to send pulses with a width of 8\un{ns} and frequency up
to 5\un{MHz}. Calibration with X-rays are performed facing $^{55}$Fe and
$^{57}$Co radioactive sources.
Preliminary tests on chips with a different number of aluminum pixels (from 1 to 9),  allowed to optimize the entire setup and characterize the first prototypes~\cite{Cruciani:2014yda}.

In the second phase of the project
we will test the light detectors by facing them to a small array of \TEO\ bolometers (2-8 detectors). The measurements will be performed at Laboratori Nazionali del Gran Sasso and in the test dilution refrigerator of CUPID 
that will be equipped with the KID readout.
Our aim is to prove that not only different KIDs on the same light detector, but also different light detectors can be coupled to the same feedline, so as to reduce substantially the wiring in view of the scale-up to a thousands light detectors.

\section{Perspectives}
Presently all the subsystems required by CALDER have been setup and
the first results obtained with a 4 aluminum pixels, $2\times2\un{cm^2}$ detector are being published~\cite{CalderFirstResult}.
By the end of 2015 we foresee the first measurement with TiN or Ti/TiN films, that are expected to bring the energy resolution of the light detectors below 20\un{eV~RMS}.
The construction and running of the bolometric \TEO\ array at Gran Sasso, instrumented with CALDER's light detectors, is expected in 2016/2017.

\section*{Acknowledgements}
This work was supported by  the European Research Council (FP7/2007-2013) under contract  CALDER no. 335359
and  by the Italian Ministry of Research (FIRB 2012) under contract no. RBFR1269SL.

\bibliographystyle{spphys.bst} 
\bibliography{../../calder}

\end{document}